\documentclass{PoS}

\def\UM{\bar{\psi}{\rm{1}}\psi}        

\newcommand{\Dlr}{\stackrel{\leftrightarrow}{D}}
\newcommand{\Dl}{\stackrel{\leftarrow}{D}}
\newcommand{\Dr}{\stackrel{\rightarrow}{D}}
\newcommand{\Op}{\mathcal{O}}
\newcommand{\OpT}{\mathcal{O}^T}
\newcommand{\preprintline}{\newline
\vspace{-4.3cm}
\rightline{\parbox{12cm}
            {\hfill \rm\small DESY 06-180, Edinburgh 2006/24, Leipzig LU-ITP 2006/16, Liverpool LTH 721}}
\vspace{1.7cm}}

\title{Operator product expansion on the lattice: analytic Wilson coefficients\preprintline}

\ShortTitle{Operator product expansion on the lattice: analytic Wilson coefficients}

\author{QCDSF collaboration: 
   M.~G\"ockeler,$^a$ R.~Horsley,$^b$ H.~Perlt,$^c$\thanks{Speaker (holger.perlt@itp.uni-leipzig.de).}
   ~P.~E.~L.~Rakow,$^d$ G.~Schierholz,$^e$
   A.~Schiller$^c$\\
\llap{$^a$}Institut f\"ur Theoretische Physik, Universit\"at Regensburg,\\
93040 Regensburg, Germany\\
\llap{$^b$}School of Physics, University of Edinburgh,\\
Edinburgh EH9 3JZ, UK\\
\llap{$^c$}Institut f\"ur Theoretische Physik, Universit\"at Leipzig,\\
04109 Leipzig, Germany\\
\llap{$^d$}Theoretical Physics Division, Department of Mathematical Sciences,\\
University of Liverpool, Liverpool L69 3BX, UK\\
\llap{$^e$}John von Neumann Institut NIC/DESY Zeuthen,\\
15738 Zeuthen Germany,\\
Deutsches Elektronen-Synchrotron DESY,\\
22603 Hamburg, Germany}

\abstract{We present first results for Wilson coefficients of operators up to 
first order in the covariant derivatives for the case of Wilson fermions. 
They are derived from the off-shell Compton scattering amplitude $\mathcal{W}_{\mu\nu}(a,p,q)$ of massless quarks with momentum $p$.
The Wilson coefficients are classified according to the transformation of the corresponding operators under the hypercubic group H(4).
We give selected examples for a special choice of the momentum transfer $q$. All Wilson coefficients are given
in closed analytic form and in an expansion in powers of $a$ up to first corrections.}

\FullConference{XXIVth International Symposium on Lattice Field Theory\\
                July 23-28, 2006\\
                Tucson, Arizona, USA}

\begin{document}

\section{Introduction}

The computation of structure functions is one of the challenging problems in hadronic physics.
It can be related to the operator product expansion (OPE) of the product of two conserved
vector currents between states of particles as nucleons or mesons. The  OPE allows
for the factorization of two length scales: the short (high energy) scale is associated to c-number expansion coefficients 
(Wilson coefficients) whereas the long (low energy) scale is put into the matrix elements of local operators~\cite{Wilson}.
In the case of deep inelastic scattering - the relevant physical process to extract structure functions - the short scale is defined by the 
large momentum transfer $Q^2=-q^2$, the long range scale is given by the hadronic mass $P^2=M_h^2$. 

In lattice simulations, structure functions are calculated via their moments $\mathcal{M}_n(q^2)$. Their generic form as
determined by the OPE can be written as
\begin{equation}
\mathcal{M}_n(q^2) = C^{(2)}_n(q/\mu)\, A_n^{(2)}(\mu) + \frac{C^{(4)}_n(q/\mu)}{q^2}\, A_n^{(4)}(\mu) + \dots ,
\label{moment1}
\end{equation}
where $A_n^{(2)}$ and $A_n^{(4)}$ are reduced nucleon matrix elements of local operators of twist-two and four, respectively.
$C_n^{(2)}$ and $C_n^{(4)}$ are the corresponding Wilson coefficients, $\mu$ is the renormalisation scale.
Usually, the Wilson coefficients are calculated perturbatively. 
On the other hand, the computation of the hadronic matrix elements $A_n^{(k)}$ is intrinsically
non-perturbative. In the last years lattice QCD has been proven to be a very promising tool for this purpose~\cite{lattQCD}. 

In order to relate
the computed quantities to physical observables one has to evaluate the Wilson coefficients and the operator matrix elements in
the same scheme. Furthermore, on the lattice operators can mix with other operators of the same quantum numbers but with lower mass dimensions.
Both the coefficients of leading operators and the expectation values of higher twist operators are plagued by 
unphysical renormalon contributions~\cite{renormalon1}. In the complete OPE sum these contributions must cancel~\cite{renormalon2} which is
not the case if one combines perturbative Wilson coefficients and non-perturbative hadronic matrix elements. Therefore, if one is
interested in contributions of higher twist operators to the moments of structure functions as given in (\ref{moment1}) 
it is best to carry out the OPE completely non-perturbatively on the lattice. This results in the following form for the moments
\begin{equation}
\mathcal{M}_n(q^2) = C^{(2)}_n(a,q)\, A_n^{(2)}(a) + \frac{C^{(4)}_n(a,q)}{q^2}\, A_n^{(4)}(a) + \dots ,
\label{moment2}
\end{equation}
with $a$ denoting the lattice spacing. However, $A_n^{(4)}(a)$ and $C^{(2)}_n(a,q)$ can be divergent due to mixing of the
twist-four operator with its twist-two counterpart, i.e. $A_n^{(4)}(a) \sim 1/a^2 $ and $C^{(2)}_n(a,q) \sim 1/(aq)^2$. So both the Wilson coefficients 
and the matrix elements must be computed non-perturbatively.

Starting point of the calculation is the OPE of the Compton scattering amplitude $\mathcal{W}_{\mu\nu}$ of a photon ($\hat{J}_\mu(q)$) off a nucleon target ($\big| N(P)\rangle$) with momentum $P$:
\begin{eqnarray}
\mathcal{W}_{\mu\nu} &=& \langle N(P) \big| \hat{J}_\mu(q)\hat{J}_\nu^\dagger(q) \big| N(P)\rangle \label{OPE1} \\
           &=& \sum_{m,n}\,C^m_{\mu\nu,\mu_1, \dots ,\mu_n}(q/\mu)\,\langle N(P)\, \big| \mathcal{O}(\mu)^m_{\mu_1, \dots ,\mu_n} \big| N(P)\rangle\, .
	   \label{OPE2}
\end{eqnarray}
The sum in (\ref{OPE1}) runs over all possible operators $\mathcal{O}^m_{\mu_1, \dots \mu_n}$ of spin $n$. The index $m$ distinguishes operators of the same spin. 
In order to have reasonable convergence properties of the OPE (so that it makes sense to truncate the expansion at a particular operator
dimension) we need small $P$, i.e. $P^2 \ll q^2$. Moreover, $\mathcal{W}_{\mu\nu}$
should not suffer from large lattice artefacts, so both $P$ and $q$ must be small compared to $1/a$.
These conditions result in the constraint
\begin{equation}
P^2 \ll q^2 \ll (2\pi/a)^2\, .
\label{constraint}
\end{equation}

In order to determine the moments of structure functions both nucleon matrix elements and Wilson coefficients must be computed.
It has turned out that it is essential to control lattice artefacts in the non-pertubative evaluation of the Wilson coefficients.
Therefore we have to satisfy the constraint (\ref{constraint}) as strictly as possible, but in addition a subtraction of lattice artefacts
based on a perturbative calculation of the Wilson coefficients has proven indispensable. In this paper we present
first analytic results for perturbative Wilson coefficients with lattice artefacts included so that $O(a^2)$ corrections can
be computed. Our work is part of a project initiated by the QCDSF collaboration~\cite{QCDSFOPE} to calculate the
moments of structure functions on the lattice with the OPE method.

\section{Calculation}

Let us define the reduced Wilson coefficients $c(q^2)$ as
\begin{equation}
C(a,q) = c(q^2)\, C_{BORN}(a,q)\, ,
\end{equation}
where $ C_{BORN}$ carries the index structure. On the lattice both the non-perturbative $C(a,q)$ and $C_{BORN}(a,q)$ have corrections of $O(a^2)$.
So writing
\begin{eqnarray}
 C_{BORN}(a,q)&=& C_{BORN}^{(0)} + (aq)^2 \,C_{BORN}^{(2)}  + \dots  \label{CBORN}\\
 C(a,q)&=& c^{(0)}(q^2)\,C_{BORN}^{(0)} + (aq)^2\, c^{(2)}(q^2)\,C_{BORN}^{(2)}  + \dots \, ,
\end{eqnarray}
and taking the ratio gives
\begin{equation}
\frac{C(a,q)}{C_{BORN}(a,q)} = c^{(0)}(q^2)+\left(c^{(2)}(q^2)- c^{(0)}(q^2)\right) \frac{C_{BORN}^{(2)}}{C_{BORN}^{(0)}} \, (aq)^2 + \dots \, ,
\label{Oacorr}
\end{equation}
which yields the desired $O(a^2)$ correction term.

Because the Wilson coefficients are independent of the target states we calculate the tree level Compton scattering amplitude 
$\mathcal{W}_{\mu\nu}(a,p,q)$ with off-shell quark states of momentum $p$. The calculation is performed in
momentum space using symbolic lattice Feynman rules.
The local operators $\mathcal{O}^m_{\mu_1, \dots ,\mu_n}$ in the expansion (\ref{OPE2}) are restricted to contain at most three covariant derivatives in agreement with the
intended numerical computations. We use Wilson fermions with free parameter $r$ leading to the following tower of local operators
\begin{itemize}
\item Unpolarised case: $\bar{\psi}{\rm{1}}\psi\,, \ \bar{\psi}\gamma_\mu\Dlr_\nu\psi\,,\ \bar{\psi}\Dlr_\mu\Dlr_\nu\psi\,,\ \bar{\psi}\gamma_\mu\Dlr_\nu\Dlr_\omega\Dlr_\rho\psi$
\item Polarised case: $\bar{\psi}\gamma_\mu\gamma_5\psi\,,\ \bar{\psi}\sigma_{\mu\nu}\Dlr_\omega\psi\,,\ \bar{\psi}\gamma_\mu\gamma_5\Dlr_\nu\Dlr_\omega\psi\,,
	\ \bar{\psi}\sigma_{\mu\nu}\Dlr_\omega\Dlr_\rho\Dlr_\lambda\psi$
\end{itemize}
with $\Dlr_\mu$ the lattice covariant left-right derivative 
\begin{equation}
  \Dlr_\mu \, = \Dr_\mu \, - \Dl_\mu 
  \label{DlDr}
\end{equation}
and $\sigma_{\mu\nu}=1/2[\gamma_\mu,\gamma_\nu]$.

The amplitude $\mathcal{W}_{\mu\nu}(a,p,q)$ is expanded in powers of $\sin^n(ap)\, (n \le 3)$. Using the relation $\Dlr_\nu \sim \sin(ap_\nu)$ we can project out
all possible operators and their respective coefficients - the corresponding Born Wilson coefficients $C_{BORN}(a,q)$. In the final step we express all
operators in terms of operators transforming irreducibly under the hypercubic group H(4)~\cite{H4}.

\section{Results}

In this section we give the results for the OPE of $\mathcal{T}_{\mu\nu}$ associated to the amplitude $\mathcal{W}_{\mu\nu}$
\begin{equation}
\mathcal{W}_{\mu\nu}(a,p,q) =  \langle p| \mathcal{T}_{\mu\nu} |p\rangle \, 
\end{equation}
calculated between off-shell states of momentum $p$ (for a review see e.g.~\cite{Manohar:1992tz}).
Even when restricted to operators with at most three covariant derivatives, the general result for the expansion of $\mathcal{T}_{\mu\nu}$  is too lengthy to be presented 
in closed form - it has 880 terms.
Except for the part without covariant derivatives we give the results only for a special choice of the momentum transfer $q$ - the maximally symmetric
case
\begin{equation}
q = (f,f,f,f) , \quad s=\sin f , \quad c=\cos f\,.
\label{maxsymm}
\end{equation}
If not shown explicitly we set $a=1$ and $r=1$. The momentum transfer (\ref{maxsymm}) is chosen to present results in shortest form, other choices are
possible just as well. 
Due to lack of space we will give in this paper the first two orders in the OPE only. The complete results 
will be published in a future paper~\cite{nextOPE}. All Wilson coefficients are determined up to an overall factor which depends on the normalisation
of the corresponding operators. We adopt the normalisation as given in~\cite{H4}.

\subsection{No covariant derivative}

Here we  present the lowest order of the operator product expansion $\mathcal{T}^{(0)}_{\mu\nu}$  in the most general form
\begin{eqnarray}
\mathcal{T}^{(0)}_{\mu\nu}(a,q)/e_\gamma^2&=& -a\,r\,\delta_{\mu\nu}\,\UM - \frac{8\,a\,r \cos(a q_\mu/2)^2}{Q^2}\,\delta_{\mu\nu}\,\UM + \nonumber\\
& & \sum_\tau\,\frac{2\,a\,r\,\cos(a q_\mu/2)^2 \cos(a q_\tau)}{Q^2}\,\delta_{\mu\nu}\,\UM + \nonumber\\
& & \frac{8\,a\,r^3\,\sin(a q_\mu/2) \sin(a q_\nu/2)}{Q^2}\,\UM - \nonumber\\
& & \sum_\tau\,\frac{2\,a\,r^3\,\cos(a q_\tau) \sin(a q_\mu/2) \sin(a q_\nu/2)}{Q^2}\,\UM + \nonumber\\
& & \frac{2\,a\,r\,\cos(a q_\mu/2) \sin(a q_\nu/2)\sin(a q_\mu)}{Q^2}\,\UM + \nonumber\\
& & \frac{2\,a\,r\,\cos(a q_\nu/2) \sin(a q_\mu/2)\sin(a q_\nu)}{Q^2}\,\UM + \nonumber\\
& & \sum_{\tau,\sigma}\,\frac{2\,i\,a\,\cos(a q_\mu/2) \cos(a q_\nu/2) \sin(a q_\sigma)}{Q^2}\,\bar{\psi}\gamma_\tau\gamma_5\epsilon_{\mu\sigma\nu\tau}\psi\,,
\end{eqnarray}
with
\begin{equation}
Q^2=Q^2(a,q) = \sum_\tau \sin(a q_\tau)^2 + r^2 \left(\sum_\tau (1-\cos(a q_\tau))   \right)^2\,.
\label{Q2}
\end{equation}
$e_\gamma^2$ denotes the quark-photon coupling.

Let us give the results for the momentum choice (\ref{maxsymm}). For the diagonal part $\mathcal{T}^{(0)}_{\mu\mu}(f)$ (choosing $\mu=1$ as an example) 
we find the decomposition
\begin{equation}
\mathcal{T}^{(0)}_{11}(f)/e_\gamma^2 = - \frac{6(3-c)(1-c)}{Q^2_f} \,\UM, \quad Q_f^2=4s^2+ 16(1-c)^2 \,.
\label{W011}
\end{equation}
At this level the Wilson coefficients can simply be read off (the notation $\tau_l^{(k)}$ for the irreducible representations
is explained in~\cite{H4}):
\vspace{0.5cm}
\\
\begin{tabular} {c l l  l}
\hline
 operator & representation & Wilson coefficient & $a$ expansion\\
 \hline
\\[-1.5ex]
$\bar{\psi}1\psi$ & $\tau^{(1)}_1$   & $-\frac{6(3-c)(1-c)}{Q^2_f}$ & $-\frac{3a}{2}(1-\frac{5}{12} (af)^2)$\\[0.8ex]
\hline
\end{tabular}
\vspace{0.5cm}
\\
Reinserting the $a$ dependence, the expansion in powers of $a$ is obtained by expanding the sine and cosine functions.
The off-diagonal part also contains contributions for polarised structure functions ($\mu=1,\nu=2$ as an example):
\begin{equation}
\mathcal{T}^{(0)}_{12}(f)/e_\gamma^2 = \frac{2(3-c)(1-c)}{Q^2_f}\UM +  \frac{i(1+c) s}{Q^2_f}(\bar{\psi}\gamma_3\gamma_5\psi - \bar{\psi}\gamma_4\gamma_5\psi)
\label{W012}
\end{equation}
The Wilson coefficients are obtained as
\vspace{0.5cm}
\\
\begin{tabular} {c l  l l}
\hline
 operator & representation & Wilson coefficient  & $a$ expansion\\
 \hline
\\[-1.5ex]
$\bar{\psi}1\psi$                  & $\tau^{(1)}_1$  & $ \frac{2(3-c)(1-c)}{Q^2_f}$    & $\frac{a}{2}(1-\frac{5}{12} (af)^2)$\\[0.5ex]
$\bar{\psi}\gamma_3\gamma_5\psi,-\bar{\psi}\gamma_4\gamma_5\psi$ & $\tau^{(4)}_4$ & $ \frac{i(1+c)s}{Q^2_f}$     & $\frac{i}{2f}(1-\frac{13}{12} (af)^2)$\\[0.8ex]
\hline
\end{tabular}
\vspace{0.5cm}
\\

\subsection{One covariant derivative}

In this case the local operators $\Op_{\mu\nu}=\bar{\psi}\gamma_\mu\Dlr_\nu\psi$ and $\OpT_{\mu\nu\omega}=\bar{\psi}\sigma_{\mu\nu}\Dlr_\omega\psi$ contribute. For the diagonal part $\mathcal{T}^{(1)}_{11}(f)$
we find four independent functions $b_i(c,s)$ which build the corresponding Wilson coefficients
\begin{eqnarray}
b_1&=& 4 i (1-c)(74-126c+63c^2-9c^3)/Q_f^4 \nonumber\\
b_2&=& -4 i (6-8c+3c^2)s^2/Q_f^4 \nonumber\\
b_3&=& 4 i (4-3c)s^2/Q_f^4 \nonumber\\
b_4&=& -4 i (1-c)(4-9c+3c^2)/Q_f^4
\end{eqnarray}
The following table presents the Wilson coefficients belonging to definite representations of operators building $\mathcal{T}^{(1)}_{11}(f)$.
\vspace{0.5cm}
\\
\begin{tabular} {c l l  l}
\hline
 operator & repr.  & Wilson coeff. & $a$ expansion\\
 \hline
\\[-1.5ex]
$\frac{1}{2}(\Op_{11}+ \Op_{22}+\Op_{33}+\Op_{44}) $                                          & $\tau^{(1)}_1$   & $\frac{1}{2}(b_1+3b_3)$       & $\frac{i}{2f^2}(1+\frac{29}{24} (af)^2)$\\[0.5ex]
$\frac{1}{2}(\Op_{11}+ \Op_{22}-\Op_{33}-\Op_{44}),\frac{1}{\sqrt{2}}(\Op_{11}-\Op_{22})$   & $\tau^{(3)}_1$   & $\frac{1}{2}(b_1-b_4)$        & $\frac{15 i \,a^2}{16}(1-\frac{4}{3} (af)^2)$\\[0.5ex]
$\frac{1}{\sqrt{2}}(\Op_{12}+\Op_{21}),\frac{1}{\sqrt{2}}(\Op_{13}+\Op_{31}),\frac{1}{\sqrt{2}}(\Op_{14}+\Op_{41})$              & $\tau^{(6)}_3$   & $\frac{1}{\sqrt{2}}(b_2+b_3)$ & $\frac{i \,a^2}{8\sqrt{2}}(1-\frac{4}{3} (af)^2)$   \\[0.5ex]
$\frac{1}{\sqrt{2}}(\Op_{23}+\Op_{32}),\frac{1}{\sqrt{2}}(\Op_{24}+\Op_{42}),\frac{1}{\sqrt{2}}(\Op_{34}+\Op_{43})$              & $\tau^{(6)}_3$   & $\sqrt{2}b_3$                 & $\frac{i}{2\sqrt{2}f^2}(1-\frac{1}{6} (af)^2)$  \\[0.5ex]
$\frac{1}{\sqrt{2}}(\Op_{12}-\Op_{21}),\frac{1}{\sqrt{2}}(\Op_{13}-\Op_{31}),\frac{1}{\sqrt{2}}(\Op_{14}-\Op_{41})$              & $\tau^{(6)}_1$   & $\frac{1}{2}(b_2-b_3)$        & $-\frac{i}{4 f^2}(1-\frac{5}{12} (af)^2)$ \\ \\[-1.2ex]
\hline
\end{tabular}
\vspace{0.5cm}
\\
For the off-diagonal part 
$\mathcal{T}^{(1)}_{12}(f)$ we have 12 combinations $b_i(c,s)$ 
\begin{eqnarray}
b_1&=& -4 i (1-c)(33-52c+24c^2-3c^3)/Q_f^4\nonumber\\
b_2&=& 4 i (7-9c+3c^2)s^2/Q_f^4\nonumber\\
b_3&=& -4 i (3-2c)s^2/Q_f^4\nonumber\\
b_4&=& 6 i (1-c)^2(1+c)(2-c)/Q_f^4\nonumber\\
b_5&=&  2 i (1-c)^2(4-9c+3c^2)/Q_f^4\nonumber\\
b_6&=& 2 i (1-c)^2(1+c)(4-3c)/Q_f^4\nonumber\\
b_7&=&  -4  (1-c)(19-18c+3c^2)s/Q_f^4\nonumber\\
b_8&=& 2  (1-c)(14-15c+3c^2)s/Q_f^4\nonumber\\
b_9&=&  12 (2-c)s^3/Q_f^4\nonumber\\
b_{10}&=& -4 s^3/Q_f^4\nonumber\\
b_{11}&=& -2  (1-c)(4-9c+3c^2)s/Q_f^4\nonumber\\
b_{12}&=& -2  (4-3c)s^3/Q_f^4
\end{eqnarray}
The corresponding Wilson coefficients are shown in the last table.
\begin{table}[!htb]
\begin{tabular} {c l l  l}
\hline
 operator & repr.  & Wilson coeff. & $a$ expansion\\
 \hline
\\[-1.5ex]
$\frac{1}{2}(\Op_{11}+ \Op_{22}+\Op_{33}+\Op_{44}) $                                                          & $\tau^{(1)}_1$   & $b_1+b_5$                          & $-\frac{i}{4f^2}(1+\frac{25}{12}(af)^2)$\\[0.5ex]
$\frac{1}{2}(\Op_{11}+ \Op_{22}-\Op_{33}-\Op_{44})$                                           & $\tau^{(3)}_1$   & $b_1-b_5$                          & $-\frac{i}{4f^2}(1+\frac{19}{12}(af)^2$)\\[0.5ex]
$\frac{1}{\sqrt{2}}(\Op_{12}+\Op_{21})$                                                       & $\tau^{(6)}_3$   & $\sqrt{2}b_2$                      & $\frac{i}{2\sqrt{2}f^2}(1-\frac{1}{6}(af)^2)$\\[0.5ex]
$\frac{1}{\sqrt{2}}(\Op_{13}+\Op_{31}),\frac{1}{\sqrt{2}}(\Op_{14}+\Op_{41})$                                           & $\tau^{(6)}_3$   & $\frac{1}{\sqrt{2}}(b_3+b_4)$      & $\frac{i}{4\sqrt{2}f^2}(1+\frac{1}{12}(af)^2)$\\[0.5ex]
$\frac{1}{\sqrt{2}}(\Op_{23}+\Op_{32}),\frac{1}{\sqrt{2}}(\Op_{24}+\Op_{42})$                                           & $\tau^{(6)}_3$   & $\frac{1}{\sqrt{2}}(b_3+b_4)$      & $\frac{i}{4\sqrt{2}f^2}(1+\frac{1}{12}(af)^2)$\\[0.5ex]
$\frac{1}{\sqrt{2}}(\Op_{34}+\Op_{43})$                                           & $\tau^{(6)}_3$   & $\sqrt{2}b_6$                      & $\frac{i a^2}{8\sqrt{2}}(1-\frac{4}{3}(af)^2)$\\[0.5ex]
$\frac{1}{\sqrt{2}}(\Op_{13}-\Op_{31}),\frac{1}{\sqrt{2}}(\Op_{14}-\Op_{41})$                                           & $\tau^{(6)}_1$   & $\frac{1}{\sqrt{2}}(b_3-b_4)$      & $\frac{i}{4\sqrt{2}f^2}(1-\frac{17}{12}(af)^2)$\\[0.5ex]
$\frac{1}{\sqrt{2}}(\Op_{23}-\Op_{32}),\frac{1}{\sqrt{2}}(\Op_{24}-\Op_{42})$                                           & $\tau^{(6)}_1$   & $\frac{1}{\sqrt{2}}(b_3-b_4)$      & $\frac{i}{4\sqrt{2}f^2}(1-\frac{17}{12}(af)^2)$\\[0.5ex]
$\sqrt{\frac{2}{3}}(\OpT_{1\{23\}}+\OpT_{2\{13\}}),\sqrt{\frac{2}{3}}(\OpT_{1\{24\}}+\OpT_{2\{14\}})$  & $\tau^{(8)}_2$   & $\sqrt{\frac{2}{3}}(2b_{10}+b_9)$  & $\frac{a}{2\sqrt{6}f}(1-\frac{4}{3}(af)^2)$\\[0.5ex]
$\sqrt{\frac{2}{3}}(\OpT_{1\{34\}}+\OpT_{3\{14\}}),-\sqrt{\frac{2}{3}}(\OpT_{2\{34\}}+\OpT_{3\{24\}})$                                & $\tau^{(8)}_2$   & $\sqrt{6}b_{12}$                & $-\frac{\sqrt{3}a}{4\sqrt{2}f}(1-\frac{4}{3}(af)^2)$\\[0.5ex]
$\sqrt{2}\OpT_{2\{13\}},\sqrt{2}\OpT_{2\{14\}}$                                         & $\tau^{(8)}_2$   & $-\sqrt{6}b_{9}$               & $-\frac{3\sqrt{3}a}{2\sqrt{2}f}(1-\frac{4}{3}(af)^2)$\\[0.5ex]
$\sqrt{2}\OpT_{3\{14\}},-\sqrt{2}\OpT_{3\{24\}}$                                         & $\tau^{(8)}_2$   & $-\sqrt{2}b_{12}$               & $\frac{a}{4\sqrt{2}f}(1-\frac{4}{3}(af)^2)$\\[0.5ex]
$\frac{1}{\sqrt{2}}(\OpT_{122}-\OpT_{133}),-\frac{1}{\sqrt{6}}(\OpT_{122}+\OpT_{133}-2\OpT_{144})$           & $\tau^{(8)}_1$   & $\sqrt{2}(b_{11}+b_7)$             & $-\frac{3a}{4\sqrt{2}f}(1-\frac{4}{3}(af)^2)$\\[0.5ex]
$\frac{1}{\sqrt{2}}(\OpT_{211}-\OpT_{233}),\frac{1}{\sqrt{2}}(\OpT_{211}+\OpT_{233}-2\OpT_{244})$          & $\tau^{(8)}_1$   & $\sqrt{2}b_{11}$                           & $\frac{a}{4\sqrt{2}f}(1-\frac{4}{3}(af)^2)$\\[0.5ex]
$\frac{1}{6}\sum_{p\in(1,2,3)}{\rm sgn}(p)\OpT_{123},\frac{1}{6}\sum_{p\in(1,2,4)}{\rm sgn}(p)\OpT_{124} $            & $\tau^{(4)}_4$   & $\sqrt{\frac{2}{3}}(b_9-b_{10})$   & $\sqrt{\frac{2}{3}}\frac{a}{f}(1-\frac{4}{3}(af)^2)$\\[0.5ex]
$\frac{1}{\sqrt{3}}(\OpT_{122}+\OpT_{133}+\OpT_{144}) $                                & $\tau^{(4)}_1$   & $\frac{4}{\sqrt{3}}(4b_{11}-2b_7)$ & $\frac{2\sqrt{3}a}{f}(1-\frac{4}{3}(af)^2)$\\ [0.5ex]
$\frac{1}{\sqrt{3}}(\OpT_{211}+\OpT_{233}+\OpT_{244}) $                             & $\tau^{(4)}_1$   & $-\frac{4}{\sqrt{3}}b_{11}$        & $-\frac{a}{2\sqrt{3}f}(1-\frac{4}{3}(af)^2)$\\ \\[-1.2ex]
\hline
\end{tabular}
\end{table}
\\

\section{Summary}

In this paper we have presented first results for perturbative Wilson coefficients in lattice QCD needed to
compute moments of structure functions. We have used Wilson fermions to calculate the off-shell Compton scattering
amplitude. The expansion in the lattice spacing $a$ up to $O(a^2)$ can be used to determine the lattice
artefacts in the numerical determination of the Wilson coefficients. As can be seen from the tables there are
Wilson coefficients which are at least of $O(a)$. They are associated to the operators $\bar{\psi}{\rm1}\psi$
and $\bar{\psi}\sigma_{\mu\nu}\Dlr_\omega\psi$ which do not occur in the continuum OPE for $\mathcal{T}_{\mu\nu}$.

\vspace{1cm}
\hspace{-0.9cm}
{\it Acknowledgements.} This work is supported by DFG under contract number FOR 465 (Forschergruppe Gitter-Hadronen-Ph\"anomenologie) and
by the EU Integrated Infrastructure Initiative Hadron Physics under contract number RII3-CT-2004-506078.


\begin{thebibliography}{99}
\bibitem{Wilson} 
K.~Wilson, \emph{Phys.~Rev.} {\bf 179} (1969) 1499.


\bibitem{lattQCD}
for a review of latest results see: K.~Orginos, \emph{PoS (LAT2006)}.

\bibitem{renormalon1}
G.~t'Hooft, \emph{The Whys of Subnuclear Physics}, Erice 1977, Plenum Press, New York; B.~Lautrup, \emph{Phys.~Lett.} {\bf B69} (1977) 109.

\bibitem{renormalon2}
M.~Beneke, \emph{Phys.~Rept.} {\bf 317} (1999) 1 [arXiv:hep-ph/9807443].

\bibitem{QCDSFOPE}
S.~Capitani et al. (QCDSF collaboration), \emph{Nucl.~Phys.~Proc.~Suppl.} {\bf 79} (1999) 173 [arXiv:hep-ph/9906320]; 
S.~Capitani et al. (QCDSF collaboration), \emph{Nucl.~Phys.~Proc.~Suppl.} {\bf 73} (1999) 288 [arXiv:hep-lat/980971].

\bibitem{H4}
M.~G\"ockeler et al. (QCDSF collaboration), \emph{Phys.~Rev.} {\bf D54} (1996) 5705 [arXiv:hep-lat/9602029]; M.~G\"ockeler, \emph{unpublished notes}.

\bibitem{Manohar:1992tz}
A.~V.~Manohar,
  arXiv:hep-ph/9204208.

\bibitem{nextOPE}
M.~G\"ockeler et al. (QCDSF collaboration), \emph{in preparation}.




\end{thebibliography}
\end{document}